# Mechanisms of Dendrites Occurrence during Crystallization: Features of the Ice Crystals Formation

Mark E. Perel'man [a,*], Galina M. Rubinstein and Vitali A. Tatartchenko [b,**]

[a]. *Racah Institute of Physics, Hebrew University, Jerusalem, 91904, Israel*

[b]. *Saint-Gobain Crystals, France.*

[*]). *E-mail*: mark_perelman@mail.ru; [**]). *E-mail*: vitali.tatartchenko@orange.fr

Dendrites formation in the course of crystallization presents very general phenomenon, which is analyzed in details via the example of ice crystals growth in deionized water. Neutral molecules of water on the surface are combined into the double electric layer (DEL) of oriented dipoles; its field reorients approaching dipoles with observable radio-emission in the range of 150 kHz. The predominant attraction of oriented dipoles to points of gradients of this field induces dendrites growth from them, e.g. formation of characteristic form of snowflakes at free movement of clusters through saturated vapor in atmosphere. The constant electric field strengthens DELs' field and the growth of dendrites. Described phenomena should appear at crystallization of various substances with dipole molecules, features of radio-emission can allow the monitoring of certain processes in atmosphere and in technological processes. Crystallization of particles without constant moments can be stimulated by DELs of another nature with attraction of virtual moments of particles to gradients of fields and corresponding dendrites formation.

Growth of ice crystals is among the most elementary and most accessible for research of processes of crystallization. The composition of frozen water is easy controlled; it is possible to observe unique processes of crystals-snowflakes formation without influence of gravitation. Therefore investigations of different processes of ice crystals formation can lead to exclusive information related to general problems of solidification. On the other hand these processes are very significant for practice of meteorology and for certain possibilities of weather prognosis.

Notice that physics of crystals arises at 1611, about four centuries ago, with the famous essay of Kepler on forms of snowflakes [1], but till now the particular, as can be seemed, physics of snow crystals contains a lot of unsolved problems. The general interest to them is connected, in particular, with the dendrite (or even fractal) structure features that lead to an extremely variety of investigations: cf. the recent reviews [2], the review of general problems of dendrites growth [3]. The common approach to crystallization problems consists in their consideration as the Stefan problem, possibly divided onto certain steps (e.g. [4]), with further artificial addition of some singularities of "fingers-type" forms on the flat background of the Laplace solution, but without physical substantiations of processes (there are certain considerations of the role of electric field, e.g. [5]).

The microscopic approach to the theory of phase transitions [6], especially to the phase transitions of the first kind, requires a revealing of alteration of electromagnetic interactions at entering of single particles into condensable substance. So, the most significant thermodynamic changing at such processes, removal of latent heat, should lead to emission of this energy at characteristic frequencies proportional to latent energy per particle or bond [7]. But we can not assert that such radiation, which carries away a basic part of energy and leads to the bonding of particles, represents the full content of transformation of latent heat: other radiative processes, possibly of much lesser energy, can exist and can play a significant role at formation of new phases.

We shall consider one such effect, the radio-emission at low frequencies range of $10^4 \div 10^6$ Hz, observed in the course of ice crystallization ([8] and references therein), and examine its role in this process. It will be shown that this effect can be responsible for the specific form of ice dendrites growing in natural conditions and that the external electric field can increase their growing (e.g. [9] and references therein). Certain strangeness of these phenomena consists in their occurrence in the deionized substance where are absent free charges that can radiate surplus energy and/or carry out transference of ions as in the habitual phenomena of electrocrystallization.

But these phenomena can not be produced without certain oscillations of charges; therefore it is necessary to pay attention to internal fields in researched substances. The only one internal proper field inevitably arising on a surface is the double electric layer (DEL). The formation and/or oscillations of DEL leads to the radio emission (such phenomenon was discovered and investigated in [10], references therein).

The quantum theory does not admit sharp (geometrical) borders between substances. The intermediate, transitive regions between them, in which a gradual transition, say, from dielectric susceptibilities of one substance or phase to another one occurs, always should exist. This basic feature can explain, in particular, the occurrence of quasi-liquid layer on surface of ice invented by Faraday [11], of so-called potential of freezing [12] and/or an existence of some analog of the hexagonal phase of liquid crystals. At interaction with electromagnetic fields such layer is represented by a set of virtual or real electromagnetic dipoles and/or currents [13]. So, on the border between liquid water and vapor and/or ice the DEL of quasi-liquid character with oriented dipoles and even their possible compositions must exist (such layer was considered in [14]).

Originally the structure and properties of DELs had been describing exclusively by ions distribution in both substances (e.g. [15]). But, later has been established that electric dipoles also can participate in their formation [16], and therefore in certain cases, e.g. in the deionized substance with dipoles, the DEL can be formed by them. The existence of such near-surface layer in ice is examined in [17], our consideration will support this representation. (Formation of DELs by particles with higher moments, as far as we know, is not investigated.)

Movement of charged particles (ions) in a constant electric field is a particular case of electrophoresis. Analogical movements of dipoles in such fields can be called dipolephoresis (this effect was suggested and practically used for formation of optical devices by directed diffusion of molecules with big dipole moments into polymer matrices [18, 19]).

But for relief of entering into the DEL the dipole should be correctly oriented (it will increase the attraction at comparatively big distances), and as the reorientation executes with acceleration, this turning should lead to electromagnetic emission of dipoles. For observability of emission the energy of reorientation must be no less than the energy of free rotation, i.e. the energy of corresponding degree of freedom:

$$W_{rot} \geq \kappa T . \tag{1}$$

The angular moment of dipole $\vec{d}$ orientation in the DELs' electric field $\vec{E}$ can be expressed and then approximated as

$$\vec{M} = \vec{r} \times \vec{p} = \int dt (\vec{d} \times \vec{E}) \approx (\vec{d} \times \vec{E})\tau \rightarrow (\vec{d} \times \vec{E})/f , \tag{2}$$

where $\tau$ is the duration of dipole inversion and can be approximated via the frequency of rotation: $\tau \rightarrow 1/f$. With such approximation the needed energy of dipoles reorientation can be written as

$$W_{rot} = \frac{\vec{M}^2}{2I} \approx \frac{d^2 E^2 \sin^2(2\pi f t)}{2 f^2 I}, \tag{3}$$

*I* is the moment of particle inertia.

From the condition (1) follows an estimation (upper limit) of emitting frequency:

$$f \leq \frac{|dE|}{2\sqrt{I\kappa T}}, \tag{4}$$

where the averaging $\sin^2\varphi \to 1/2$ is taken into account.

For molecules $H_2O$ the dipole moment *d* = 1.8 D and the moment of inertia $I \sim 10^{-40}$ g·cm$^2$ are generally accepted. (Note that these parameters can be précised [20], some new experiments lead to higher dipole moment of water molecules in liquid state, till 2.65 D [21], cf. also [22], but we shall use the common data and did not take into account features of water clusters [23].) The DELs' field strength $E \sim 0.1$ V/cm = $(1/3) \cdot 10^{-3}$ CGSE/cm and the estimation (4) at T = 300 K gives:

$$f \leq 150 \text{ kHz} \tag{5}$$

that corresponds, in general, to the experimental data [8]. The most part of energy of rotation (3) remains in substance, the radiated energy is much less energy of the corresponding degree of freedom, $2\pi\hbar f \ll \kappa T$, and consequently in thermodynamic parameters of system is practically not reflected. Notice that as the dielectric permeability of water $\varepsilon \sim 3$ in the examined frequencies range, interaction with surrounding particles does not essentially change these estimations.

On the surface of liquid water at vapor condensation an analogical DEL, a quasi-polarized layer, systems of oriented molecules, also must be formed [10] and the emission of frequencies $f_1 \leq 130$ kHz may be waiting. This phenomenon can describe radio noises at surfs zones, at occurrence of snow avalanches, certain noises arising with some weather phenomena, etc.

Let us notice that radiation of dipoles with their reorientation was examined by Planck [24] and just it had led to the Fokker-Planck equations. In our case, however, the limiting to more simple estimations for the initial investigation seems sufficient.

The force acting on diffusant molecules is determined as

$$\vec{F} = -\vec{\nabla}(\vec{d} \cdot \vec{E}), \tag{6}$$

i.e. at the case of constant magnitude of dipole moment and at $\vec{E}(\vec{r}) = const$, close to a flat surface this force leads only on the dipole reorientation. But in immediate proximity to singular points, e.g. to corners of ice cluster and to tops of its dendrites the electrostatic field becomes non-uniform and can be represented as a field of one big dipole $d_{DEL}$ (probably on a background of constant field of other parts of crystal). Angular dependence of its interaction with flying molecules is complicating:

$$F \sim \frac{d \cdot d_{DEL}}{r^4} f(\vartheta, \varphi). \tag{7}$$

But without further consideration it must be underlined that these forces are attractive for well oriented dipoles onto bigger distances than intermolecular forces. They collect molecules to corners of clusters and thus impoverish their lateral sides.

So, at a qualitative level, at least, this consideration leads to such conclusions: if the crystal nucleus was flat, it will grow as a stellar flat figure, the polyhedron can grow into a three-dimensional acicular form, the cylinder should grow along a length as the DEL field is non-uniform only close to its ends. Note that such representation corresponds to the common determination of duration of diffusion $\tau_{diff} \sim R^2/D$, where $R$ is a characteristic size of considered tops and $D$ is a scalar diffusion constant with taking into account properties of DEL.

For a common estimation of the role of this effect at crystal growing the generalized form of the Fick second law (the Smoluchowski equation) must be considered:

$$\frac{\partial c(\vec{r})}{\partial t} = D\vec{\nabla} \cdot \left( \vec{\nabla} c(\vec{r}) + \frac{c(\vec{r})}{\kappa T} \vec{F} \right), \tag{8}$$

where $c(\vec{r})$ is the density of solute or vapor, and at each point on the surface this density depends on surface curvature owing to the Gibbs–Thomson effect [2]. So it can be proposed that $c(\vec{r}) = c_{sat}(1 + \alpha/R\kappa T)$, where $\alpha$ describes features of DEL, i.e. the density of surface energy.

The equation (8) also demonstrates that the constant external electric field does not lead to direct attraction of dipoles to flat surface, but can only turn them. Near to corners of cluster and tops of dendrite (or near casual micro-heterogeneity on a smooth facet of crystal), where this field will be effectively non-homogeneous, its gradient can play a sufficient role. It means, in accordance with observations [8], that the additional electric field should enlarge the growth of dendrite structures in definite direction.

If the equation (8) will be written separately for dipoles with orientation to and opposite field, $c_+$ and $c_-$, then for their difference $\Delta c = c_+ - c_-$ can be written the approximation:

$$\frac{\Delta c}{\tau} \sim D\frac{\Delta c}{r^2} \pm D\frac{c}{r^2 \kappa T} 2|\vec{d} \cdot \vec{E}|. \tag{9}$$

It leads to such semi-qualitative estimation:

$$\frac{\Delta c}{c} \sim \pm \frac{dE}{\kappa T} \cdot \frac{2D}{D - fr^2}. \tag{10}$$

Near to dendrite top the first term on the right in (9), which corresponds to diffusion without participation of a field, can be omitted. It shows the accumulation of correctly oriented dipoles at points of growth and, accordingly, their decreasing in other areas. Thus the external field can accelerate the entering of dipoles into definite points of condensate. On the other hand as at usual conditions for water vapor in clouds $\Delta c$ is maximal very close to growing crystals only, the value of (10) must be proportional to intensity of emission at the duration of crystals grows. In the example considered above, $dE/\kappa T$ is of the order of $10^{-8}$ and it conforms to smallness of radiation emission fixed in [8] (as far as we know, the detection of radio emission in external fields is not yet investigated).

Here must be underlined that the requirements of sufficiently big values of $\Delta c$ at points of dendrites growth forbid their appearance at places of big gathering of singularities, e.g. at singular facets of crystal. The levelling of singularities leads to levelling of density of flying particles, i.e. to normal processes of regular crystallization.

Let's consider the denominator of this expression. At frequencies (5) and with the diffusion coefficient $D = 6 \cdot 10^{-5}$ m$^2$/s for the ice, the radio emission at crystal growing in vapor phase can be maximized, if dipoles will be reorienting at distances of the order of 2 mcm from surface.

\* \* \*

Can the considered features be peculiar to substances atoms/molecules of which do not possess dipole moments? If on border of medium the DEL is present, it will induce dipole moment $\vec{d} = \alpha \vec{E}$ in approaching particles, where $\alpha$ is the polarizability of particle, scalar in simple case. Force of its interaction with DEL is determined by expression (6), i.e.

$$\vec{F} = -\vec{\nabla}(\vec{d} \cdot \vec{E}) \to -\alpha \vec{\nabla} E^2. \tag{11}$$

Hence, such particle will be mainly attracted to points of field gradients and therefore it will promote growth of dendrites. And as an intensity of radiation of a dipole is proportional to the second derivative of the moment (spatial derivatives are expressible through derivatives on time) in these points of fields heterogeneity a low-frequency radio-emission should be generated. But this radiation would be of spread-spectrum and therefore the consideration of concrete cases is demanded.

Thus, it is possible to believe, that the basic distinction between formation of dendrites by particles with constant dipole moments and without them will consist in the spectra of an accompanying radio-emission.

\* \* \*

Let us enumerate the results of investigation, certain their generalizations and perspectives.

1. Near-surface layer, at least, of frozen water, i.e. of ice contains the ordering dipole molecules. The existence of this DEL is proved by considered features of crystallization. Notice that this phenomenon explains the high surface conductivity and related phenomena that can be assigned to the high concentration of defects (thin ice layer on the metal surfaces can has another structure [25]).

2. The crystallization and condensation of water can be considered as entering of single molecules through the DEL onto the surface of condensate.

3. This process should be accelerated at sufficiently isolated points of non-uniformity of field of DEL with a due excess of attracted free particles. The acceleration would be observable near to corners of water clusters and to tops of dendrites; this peculiarity can explain complicating and surprising variety of snowflakes with their dendrites or even their fractal forms.

4. On a singular facet of crystal a lot of singular points are leading to levelling of densities of attracting particles and to absence of conditions for dendrite growth. On the contrary, single isolated fluctuations on non-singular facets can attract surplus condensable particles and induce dendrites formation.

5. Reorientation of water dipoles in the course of their approaching to the surface DEL leads to the characteristic radio-emission.

6. External constant electric field, that is added to the field of DEL, can accelerate these processes in definite directions.

7. The attraction of electromagnetic moments to singularities of DELs on surfaces can be considered as sufficient general feature of crystallization of substances with own dipole moments of molecules. The radiation at moments of dipoles reorientation can be used for monitoring these processes. This phenomenon can be used also for monitoring certain weather processes, status of snow and ice covers and so on. (These processes are usually considered by thermodynamic approaches only [26].)

8. These representations can be evidently generalized. So, on each surface of condensed substances of almost every type the DELs must be inevitably formed (exclusions can be pure neutral substances like $CO_2$, noble gases, etc), and in atoms and molecules without constant moments they nevertheless can be virtually actuated. Therefore the crystallization with attraction to near-surface DELs can be analogical to processes considered above (without reorientation of dipoles and considered radio-emission acts, of course). Thus the attraction of flying particles will be stronger near to singularities of DEL's field and will lead to dendrites formation.

9. Slightly another example represents the electrocrystallization: singularities of field can be induced by different fluctuations and attraction to their points can be described by the usual Coulomb forces.

# REFERENCES

[1]. J. Kepler. *On the Six-Cornered Snowflakes.* English Transl.: Oxford, Clarendon, 1966; Russian Transl.: Moscow, Nauka, 1972.

[2]. V. F. Petrenko and R. W. Whitworth, *Physics of Ice*. Oxford: Oxford Univ. Press, 1999;

K. G. Libbrecht. Rep. Prog. Phys. **68** (2005) 855; Phys. World, **21**(1) (2008) 19-23.

[3]. E. A. Brener and V. I. Melnikov. Adv. Phys., **40** (1991) 53.

[4]. Th. Kim, M. Henson and M.C. Lin. In: http://www.eg.org/EG/DL/WS/SCA/SCA04/305-314

[5]. N. M. Zubarev. Phys. Lett. A, **243** (1998) 128.

[6]. M. E. Perel'man. Phil. Mag., **87** (2007) 3129.

[7]. M. E. Perel'man and V. A. Tatartchenko. Phys. Lett. A, doi:10.1016/j.physleta.2007.11.056 .

M. E. Perel'man and V. A. Tatartchenko. In arXiv: 0711.3570v2.

[8]. A. A. Shibkov et al. J. Cryst. Growth, **236** (2002) 434; Cryst. Rep. **51** (2006) 96.

[9]. K. G. Libbrecht, T. Crosby and M. Swanson. J. Cryst. Growth **240** (2002) 241.

[10]. N. G. Khatiashvili, M. E. Perel'man. Phys. Earth & Plan. Int., **57** (1989) 177.

[11]. M. Faraday. Phil. Mag., **17** (1859) 162.

[12]. L. B. Loeb. *Static Electrification*. Berlin, Springer, 1958.

[13]. M. E. Perel'man. Phys. Lett. A, **370** (2007) 528.

[14]. N. H. Fletcher. Phil. Mag. B, **66** (1992) 109.

[15]. R. Chang. *Electrochemistry*, 7th Ed., Mc Graw Hill, (2002)

[16]. J. O'M. Bockris, M. A.V. Devanathan and K. Müller, Proc. Roy. Soc. A, **274** (1963) 55.

[17]. I. A. Ryzhkin and V. F. Petrenko. Phys. Rev. B, **65** (2002) 012205.

[18]. G. M. Rubinstein and V. S. Chagulov. In: *Polymer Optical Materials*. Acad. Sci. USSR: Chernogolovka, 1989, 112-126 (In Russian).

[19]. N. G. Lekishvili et al. *Polymers and Polymeric Materials for Fiber and Gradient Optics (New Concepts in Polymer Science)*. Brill Academic Pub: 2002, p.127.

[20]. R. Moro, et al. Phys. Rev. A **75**, (2007) 013415.

[21]. Y. Tu and A. Laaksonen. Chem. Phys. Lett., **329** (2000) 283.

[22]. R. Bukowski, et al. Science, **315** (2007) 1249.

[23]. F. N. Keutsch and R. J. Saykally. PNAS, **98** (2001) 10533.

[24]. M. Planck. Sitzungsber. Akad. Wiss. Berlin, 1915, S.512.

[25]. M. Mehlhorn and K. Morgenstern. Phys. Rev. Lett., **99** (2007) 246101.

[26]. Y. S. Djikaev, et al. J. Phys. Chem. A, **106** (2002) 10247